# ppOpen-AT : A Directive-base Auto-tuning Language[*]

Takahiro Katagiri

Information Technology Center, Nagoya University

katagiri@cc.nagoya-u.ac.jp

August 29, 2024

## Abstract

ppOpen-AT is a domain-specific language designed to ease the workload for developers creating libraries with auto-tuning (AT) capabilities. It consists of a set of directives that allow for the automatic generation of code necessary for AT by placing annotations in the source program. This approach significantly reduces the effort required by numerical library developers. This technical report details the implementation of the AT software and its extended functions, and provides an explanation of the internal specifications of ppOpen-AT.

## 1. Introduction

ppOpen-AT is a domain-specific language that provides a set of directives designed to reduce the workload for developers of numerical libraries with auto-tuning (AT) features. This scripting language facilitates the automatic generation of code necessary for AT by allowing developers to place annotations in the source program, thereby minimizing their workload.

While the concept of a library with auto-tuning features is broadly applicable across various processes, the current specification of ppOpen-AT is specialized for parallel-computation processing.

Numerous studies have explored the application of AT to numerical libraries. Notable examples of numerical libraries that utilize AT include FFTW [1], ATLAS [2], ABCLib_DRSSED [3], ppOpen-HPC [4], an FDM application [5], and FFTE [6].

---

[*] This technical report serves as an updated edition of the ppOpen-AT theoretical manual (version 1.0.0), originally published in March 2016.

Additionally, when considering AT systems integrated with computer systems, such as language processing systems, several notable examples include SPL [8], Loop Transformation Recipes [9], ADAPT [10], OpenTunar [11], Xevolver [12], and ppOpen-AT, which is designed for a hybrid MPI (Message Passing Interface) and OpenMP environment [13]. The ppOpen-AT is built on the FIBER framework [14][15], which is an AT framework. Its directive-based AT language development relies on a language specifically designed for AT [16].

This technical report is structured as follows: Section 2 describes the target computer language of ppOpen-AT. Section 3 outlines the installation process of ppOpen-AT. Section 4 details the implementation of an auto-tuning library using ppOpen-AT. Section 5 discusses the extended functions of ppOpen-AT, and Section 6 provides an in-depth explanation of the internal specifications. Finally, the report concludes with a summary of the key points.

## 2. Target Computer Languages

ppOpen-AT is a domain-specific computer language, consisting of a set of directives, designed to streamline the development of parallel-computation libraries†. Given its focus on efficiency in this context, ppOpen-AT must interface with languages that are well-suited for numerical computation and parallel processing. The primary purpose of ppOpen-AT is to generate executable program code.

Accordingly, it is intended that ppOpen-AT will be used to produce code that operates in environments where Fortran90 or C are available for numerical calculations, and Message Passing Interface (MPI) and/or OpenMP are used for parallel processing.

### 2.2 Sequence of Library Development and Use

Fig. 1 illustrates the sequence of library development using ppOpen-AT, and the use of mathematics libraries with auto-tuning features by the user.

---

† The language is not limited to library development. It can also be used to optimize any user program.

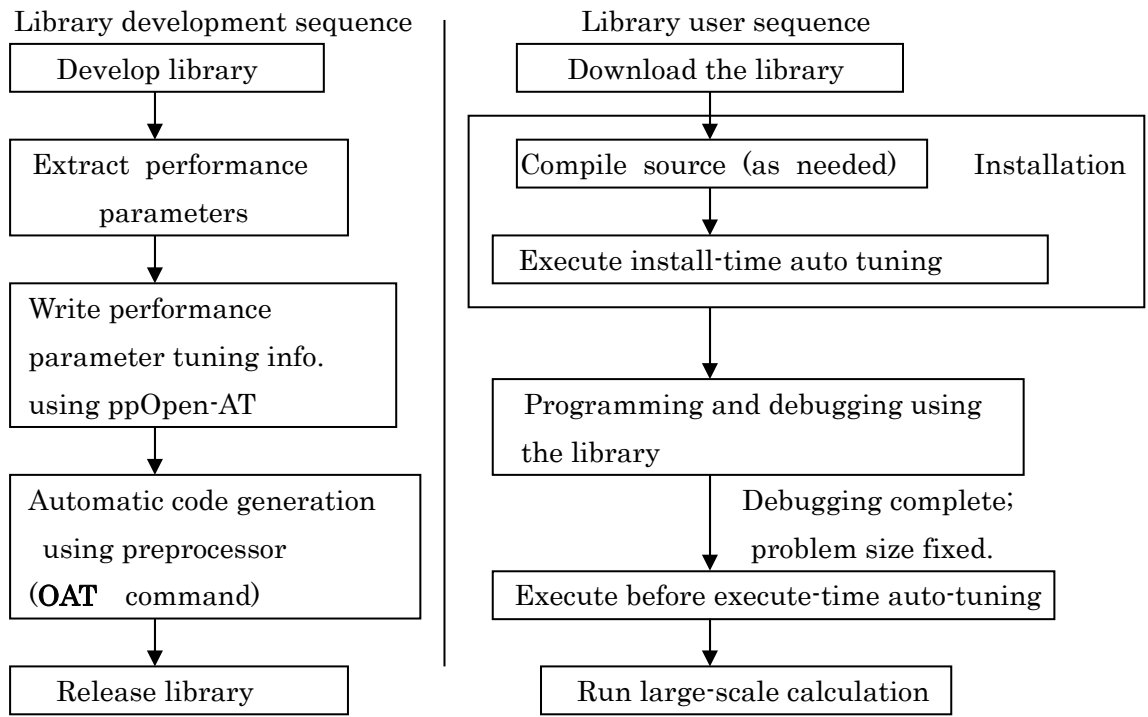

Figure 1. Sequence of Actions Required of Library Developers and Users Using ppOpen-AT.

## 2.3 Features of Programming with ppOpen-AT

ppOpen-AT was designed with the following four points in mind.

1. <u>It is an auto-tuning specification function specialized for numerical calculation</u>

    Auto-tuning features optimized for numerical calculation are provided. They facilitate the creation of numerical calculation libraries with auto-tuning features.

2. <u>It is a tuning specification method whereby the library developer inserts instructions into the program in the form of annotations.</u>

    The addition of auto-tuning features is performed in the form of annotations by the library developer. Consequently, this makes it easy to specify auto tuning, while not interfering with the compilation of the original code.

3. <u>A preprocessor that parses the library developer's annotations and automatically generates code is provided.</u>

    This feature parses the commands written by the library developer via annotations, and automatically generates code with added auto-tuning processing. This allows the library developer to manage the library source code itself, with an auto-tuning framework added on automatically. It is also possible to understand what processes have been added by viewing the automatically generated source code. This makes the processing highly transparent and worry-free.

4. Underline: The concept that two types of parameters affect performance: performance parameters (PP) and basic parameters (BP) is adopted.

The adoption of the concept of performance parameters and basic parameters improves the outlook for auto-tuning processing. It also makes it easy for library developers to communicate their intentions to the system.

## 2.4 The ppOpen-AT Software Architecture

The ppOpen-AT Application Programming Interface (API) consists of the following three elements:

- Specifiers defining auto-tuning targets, and subtype specifiers defining auto-tuning process details
- Environmental parameters defining auto-tuning information (system parameters)
- Run-time routines that supplement auto tuning

Fig. 2 illustrates the relationship between these elements.

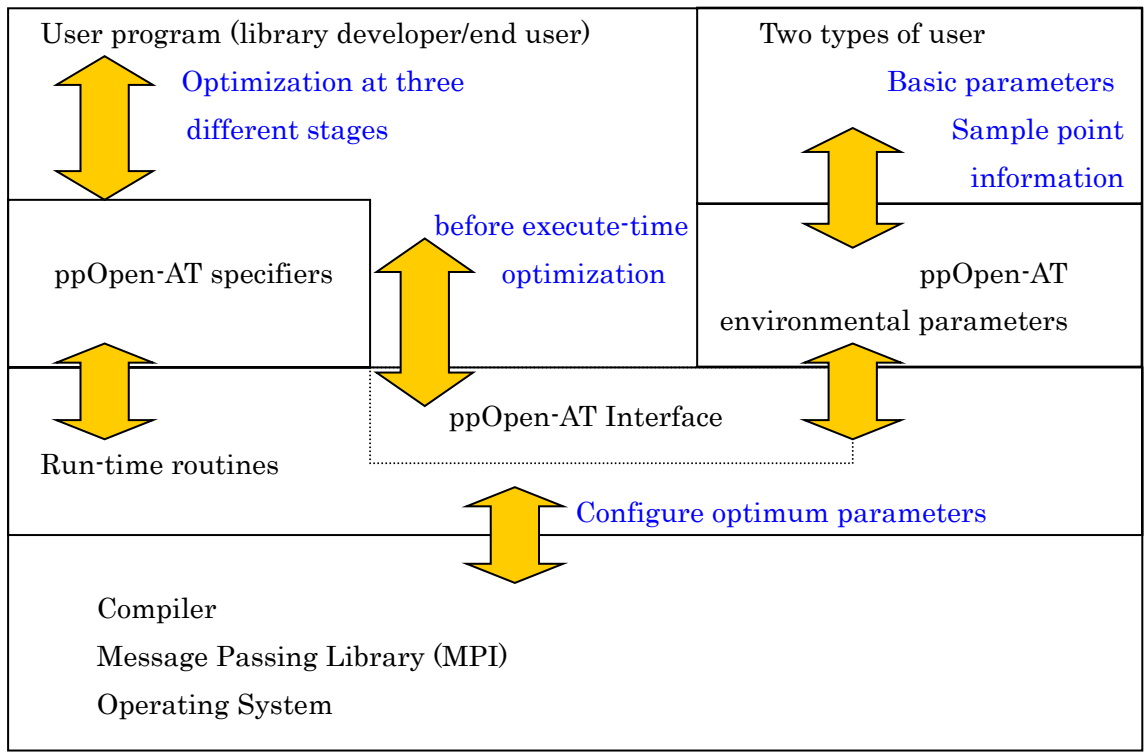

Figure 2. The ppOpen-AT Software Architecture

# 3. Getting Started

## 3.1 Auto-Tuning Features Envisioned by the System

The auto-tuning features provided by ppOpen-AT can be classified into the following three categories:

1. **Install-time auto-tuning**

   This feature automatically tunes performance parameters that can be determined when the library is installed.

2. **Before execute-time auto-tuning**

   This features auto-tunes performance parameters after the end user has fixed the problem size and other basic parameters.

3. **Run-time auto-tuning**

   This feature automatically tunes performance parameters that can be determined when the library is actually called from within a program.

These auto-tuning features are implemented as subroutines based on the Framework of Install-time, Before Execute-time and Run-time optimization feature (FIBER; patent application filed January 2003.) software architecture, an auto-tuning library architecture proposed by the ABCLibScript project, a precursor to the ppOpen-AT project.

Fig. 3 shows an overview of FIBER. With the exception of the auto-modeling routine shown in Fig. 3, ppOpen-AT is a scripting language (set of directives) developed to provide the three above-mentioned auto-tuning features.

Fig. 4 illustrates the parameter information hierarchy. In other words, parameters determined by the install-time auto-tuning routine can be referenced by the before execute-time auto tuning and run-time auto tuning routines. However, parameters determined by the before execute-time auto-tuning routine can only be referenced by the run-time auto-tuning routine. Finally, parameters determined by the run-time routine cannot be referenced by any other routine except the run-time routine[‡].

---

‡ An exception, however, is in the FIBER feedback model, where it is possible for the before execute-time routine to reference parameters optimized by the run-time routine, and optimize them.

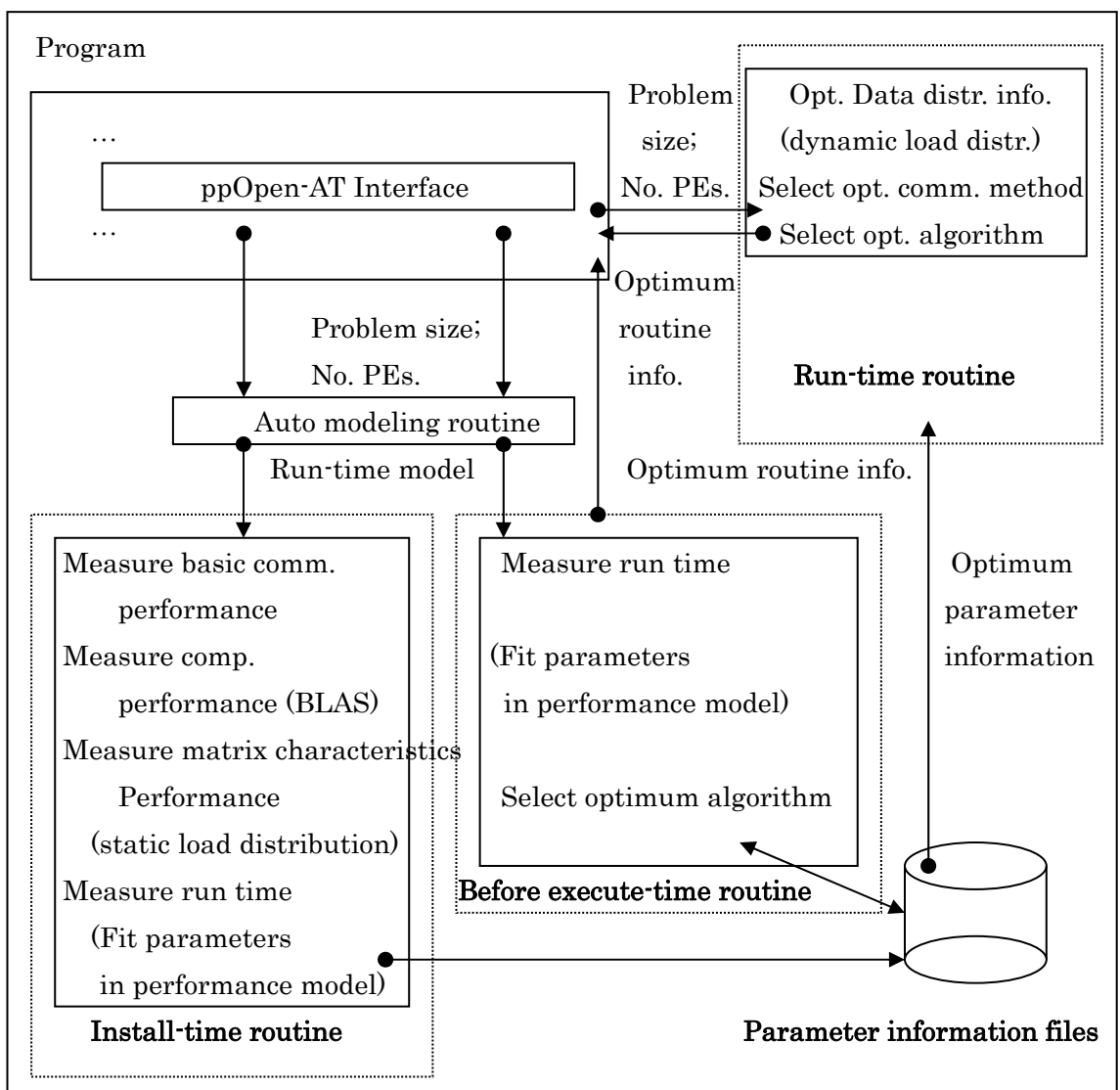

Figure 3. Structural Diagram of the Auto Tuning Library in the OAT Project
(FIBER Software Architecture)

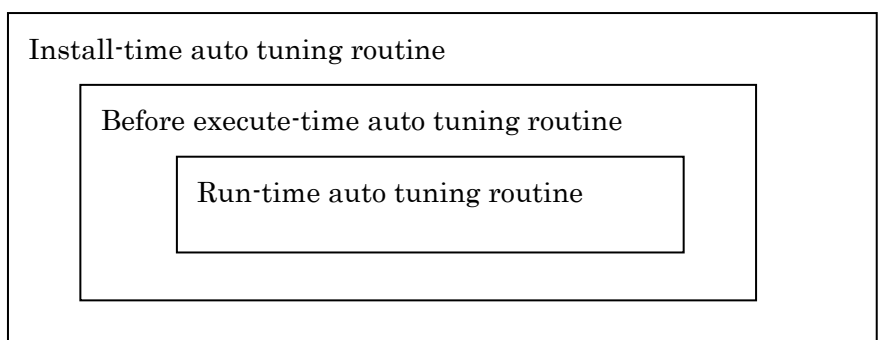

Figure 4. Hierarchy of Parameter Information Referencing

## 3.2 Execution Priority for ppOpen-AT's Auto Tuning

OAT executes auto-tuning in the following order.

1. Install-time auto tuning routine
2. Before execute-time auto tuning routine
3. Run-time auto tuning routine

Please note that if the execution sequence deviates from the specified order, an error code will be generated, and the auto-tuning process will halt.

## 3.3 Parameter Types

FIBER tunes the following two types of parameters in libraries, subroutines, and in some programs developed by users (library developers and end users).

1. **Basic parameters (*BP*)**

   These parameters must be set when the end user uses the library. Some examples of BPs include matrix size and number of processors.

2. **Performance parameters (*PP*)**

   These parameters are not absolutely necessary for the end user to utilize the library, but do impact performance. One example is the number of unrolling levels. The user guarantees that the optimum values for performance information parameters can be found if the basic parameters are set.

   ppOpen-AT is a scripting language, consisting of a set of directives, developed to simplify the process for library developers to specify FIBER's basic and performance parameters. As such, the ppOpen-AT notation described below serves as a scripting language for declaring, defining, and specifying both types of parameters.

## 3.4 Notation

### 3.4.1 Overview

With ppOpen-AT, auto-tuning features are achieved by coding processes in the form of annotations in the source program. Specifically, lines beginning with

```
!OAT$
```

are considered to be instructions to ppOpen-AT.

Briefly, the notational conventions are as follows:

```
 !OAT$  <auto-tuning type> <feature name> [(target parameter)] region  start
〔 !OAT$  <feature details> 〔 sub  region  start 〕 〕
        Program targeted for processing
〔 !OAT$  <feature details> 〔 sub  region  end〕 〕
 !OAT$  <auto-tuning type> <feature name> [(target parameter)] region  end
```

The <auto-tuning type> and <feature name> items above are called **specifiers**. The <feature details> items above are called **subtype specifiers**.

Finally, the processing surrounded by **!OAT$ region start** … **!OAT$ region end** is called the **tuning region** or **AT region**.

### 3.4.2 Specifiers

The ppOpen-AT specifiers are described in detail below.

**List of specifiers**

```
<auto-tuning type>::= (install | static | dynamic | <formula>)
        Install: Specifies install-time auto tuning
        static: Specifies before execute-time auto tuning
        dynamic: Specifies run-time auto tuning
        <formula>::= Indicates a formula conforming to Fortran90 syntax
<feature name>::= (define | variable | select | unroll)
        define: Specifies that this process sets a parameter
        variable: Specifies that this is a variable parameter
        select: Specifies that this process selects from multiple options
        unroll: Specifies that the following process performs loop unrolling
```

### 3.4.3 Subtype Specifiers

The specifier <feature name> may require a further annotation. The subtype specifier is used to encode this. The following <feature details> subtype specifiers are available.

```
<feature details>::= ( name | parameter | select | according |
                        varied | fitting | number | prepro | postpro )
```

The <feature details> subtype specifiers that can be used depend on the specifier. This is shown below.

| Specifier | Subtype specifiers available |
| --- | --- |
| define | name, parameter, number, prepro, postpro |
| variable | name, parameter, varied, fitting, number, prepro, postpro |
| select | name, parameter, select, according, number, prepro, postpro |
| unroll | name, parameter, varied, fitting, number, prepro, postpro |

Details about each subtype specifier are shown below.

List of subtype specifiers

- **name** \<string\>: Denotes the name of the tuning region
   (available for all functions)
- **parameter** ( \<attribute specification\> \[\<attribute specification\> \,…]  )
  Specify a parameter to output to a performance characteristics file, or input from
  a performance characteristics file

  > \<attribute specification\>::=[ **in**  |  **out**  |  **bp** ]
  >    **in**:   Input parameter (defined and referenced externally)
  >    **out**:   Output parameter (defined in this tuning region)
  >    **bp**:   Basic parameter
  >    (Available for all functions)

- **select  sub  region** ( **start** | **end** ):   Specifies that this is a selection procedure
   (For the function-name select specification.)
- **according** (\<conditional expression\> | **estimated** ): Specifies a selection procedure
  based on the standard specified below

  > \<conditional expression\>::=
  >    [ ( **min**(\) | **condition** (\<condition\>) )  \<connector\>]
  >    \<connector\>::=[ **.and.** | **.or.** ] \<conditional expression\>

  > **estimated**  \<mathematical expression\>:   Specifies that the optimum process
  >   should be selected and run, based on the  user-defined cost
  >   (mathematical expression) associated with the selections.

   (For the function-name select specification. )
- **Varied**  (\[, \] )  **from**  X  **to**  Y
  Specifies the range of variation of the specified parameter (from X to Y)
   (For the function-name variable/unroll specification.)
- **Fitting**  \<method\> **sampled** \<scope\>: Specifies the method to use for
   inferring parameters
  \<method\>::=
         [ **least-squares** \<order\> | **dspline** |
             **user-defined** \<mathematical expression\> | **auto** ]
     **least-squares**: Specifies the parameter must be inferred by the least squares
        method via a polynomial expression.
      \<order\>   Sets the order of the polynomial expression.
     **dspline**: Specifies the parameter must be inferred by discrete spline.
     This method was developed by the Tanaka Laboratory, Kogakuin University,.

> **user-defined**: Infer using the least squares method, using the mathematical
>      expression specified by the user.
> **auto**: Allow the system to infer the parameter.
>
> <scope>::= [ <number> | **auto** ] Specifies the scope required for
>      parameter inference. Note that this can be omitted when <method> = auto
>      <number>: Specifies the concrete numerical value of the parameter
>      **auto**: Set the parameter's sampling interval automatically.

If the fitting subtype specifier is omitted, the optimum parameter is determined by measuring the entire range specified by the varied subtype specifier (i.e. exhaustive search).

(For the function-name variable/unroll specification.)

- **number** <number>: Used to specify the order in which to process the tuning regions. If this is omitted, the regions are processed first to last. In the case of nested specifiers, a number can only be assigned to the outermost specifier (Available for all functions.)
- **prepro sub region** ( **start** | **end** ): Specifies processing to apply before calling the tuning region (Available for all functions)
- **postpro sub region** ( **start** | **end** ): Specifies processing to be applied after calling the tuning region (Available for all functions.)
- **debug** (, [, ...]): Specifies that the variable is to be shown in the debug display when the tuning region is executed

(Available for all functions.)

> ::=[ **bp** | **pp** | **any** ]
>      **bp**: Display basic parameter information
>      **pp**: Display performance parameter information
>      **Any**: Display information about the specified parameter

## Sample Program 1

Perform auto tuning upon installation to unroll a matrix product loop from 1 to 16 levels. The parameters are inferred by means of the least squares method, using a fifth-order polynomial equation. Additionally, set the sample points at 1-5, 8, and 16. Specify the performance parameter (number of unrolling levels) for debug display.

```
!OAT$   install   unroll   region   start
!OAT$   name   MyMatMul
!OAT$   varied   (i, j)   from 1 to 16
!OAT$   fitting   least-squares   5   sampled   (1-5, 8, 16)
!OAT$   debug   (pp)
do i=1, n
  do j=1, n
    do k=1,n
        A(i, j)   =   A(i, j)   +   B(i, k)   *   C(k, j)
    enddo
  enddo
enddo
!OAT$   install   unroll   (i, j)   region   end
```

# 4. Implementing an Auto Tuning Library via ppOpen-AT

## 4.1 The Initialization Interface

ppOpen-AT provides an interface for executing auto tuning. The user can specify that auto tuning is desired by means of this interface.

Specifically, ppOpen-AT provides the following interface for auto tuning.

**OAT_ATexec** ( OAT_ATkinds,   OAT_ATroutines )

The OAT_ATexec procedure performs the auto tuning specified by OAT_ATkinds within the auto-tuning region specified by OAT_ATroutines.

The OAT_ATkinds parameter is used to specify the type of auto-tuning to perform. One of the following four constants defined in the **OAT.h** header file can be specified.

- **OAT_INSTALL**:   Install-time auto tuning
- **OAT_STATIC**:   Before execute-time auto tuning
- **OAT_DYNAMIC**:   Run-time auto tuning
- **OAT_ALL**:   All auto-tuning

The OAT_ATroutines parameter specifies which tuning region to perform the processing on. This parameter can be declared by the user, using the type OAT_ATname defined in the header file OAT.h, or it can be specified using a global variable

common-defined in OAT.h:

- **OAT_AllRoutines**:　For all routines
- **OAT_InstllRoutines**:　For install-time auto tuning routines
- **OAT_StaticRoutines**:　For before execute-time auto tuning routines
- **OAT_DynamicRoutines**:　For run-time auto tuning routines

　The OAT_ATset subroutine substitutes auto-tuning processor information.

　Please note that during install-time auto-tuning and before execute-time auto-tuning, auto-tuning occurs when OAT_ATexec is called. However, during run-time auto-tuning, only the execution is configured, and the actual auto-tuning takes place when the relevant section is invoked.

**Sample Usage**

!OAT$ call OAT_ATexec ( OAT_INSTALL,　OAT_InstallRoutines )

Performs install-time optimization.

---

**OAT_ATset** ( OAT_ATkinds,　OAT_ATroutines )

---

　The OAT_ATset procedure performs the procedures configured in OAT_ATroutines on the tuning region specified by the OAT_ATkind directive.

　The OAT_ATkinds parameter is used to specify the type of auto-tuning to perform. One of the following four constants defined in the OAT.h header file can be specified.

- **OAT_INSTALL**:　Install-time auto tuning
- **OAT_STATIC**:　Before execute-time auto tuning
- **OAT_DYNAMIC**:　Run-time auto tuning
- **OAT_ALL**:　All auto-tuning

　The name of the target tuning region is substituted in the OAT_ATroutines parameter.

　Note that the following routine, OAT_ATdel, is used to delete specific routine names.

---

**OAT_ATdel** ( OAT_ATroutines, DelName)

---

　The OAT_ATdel procedure deletes the tuning region with the name specified by DelName, from the parameter containing the tuning region name information specified by means of OAT_ATroutines.

In the DelName parameter, specify the name of the tuning region that you wish to delete. Please note that the tuning region name is included for each tuning region in the annotation format (see details below).

**Sample Usage**

!OAT$ call OAT_ATdel(OAT_InstallRoutines, "MyMatMul")

Deletes the MyMatMul tuning region from the candidates for install-time optimization.

## 4.2 Overview of Implementation Methods

This section describes each function, using concrete examples.

A library developer's subroutine, EigenSolver, can be written in the following sequence, using the ppOpen-AT interface in the library developer's subroutine, foo, which performs auto tuning.

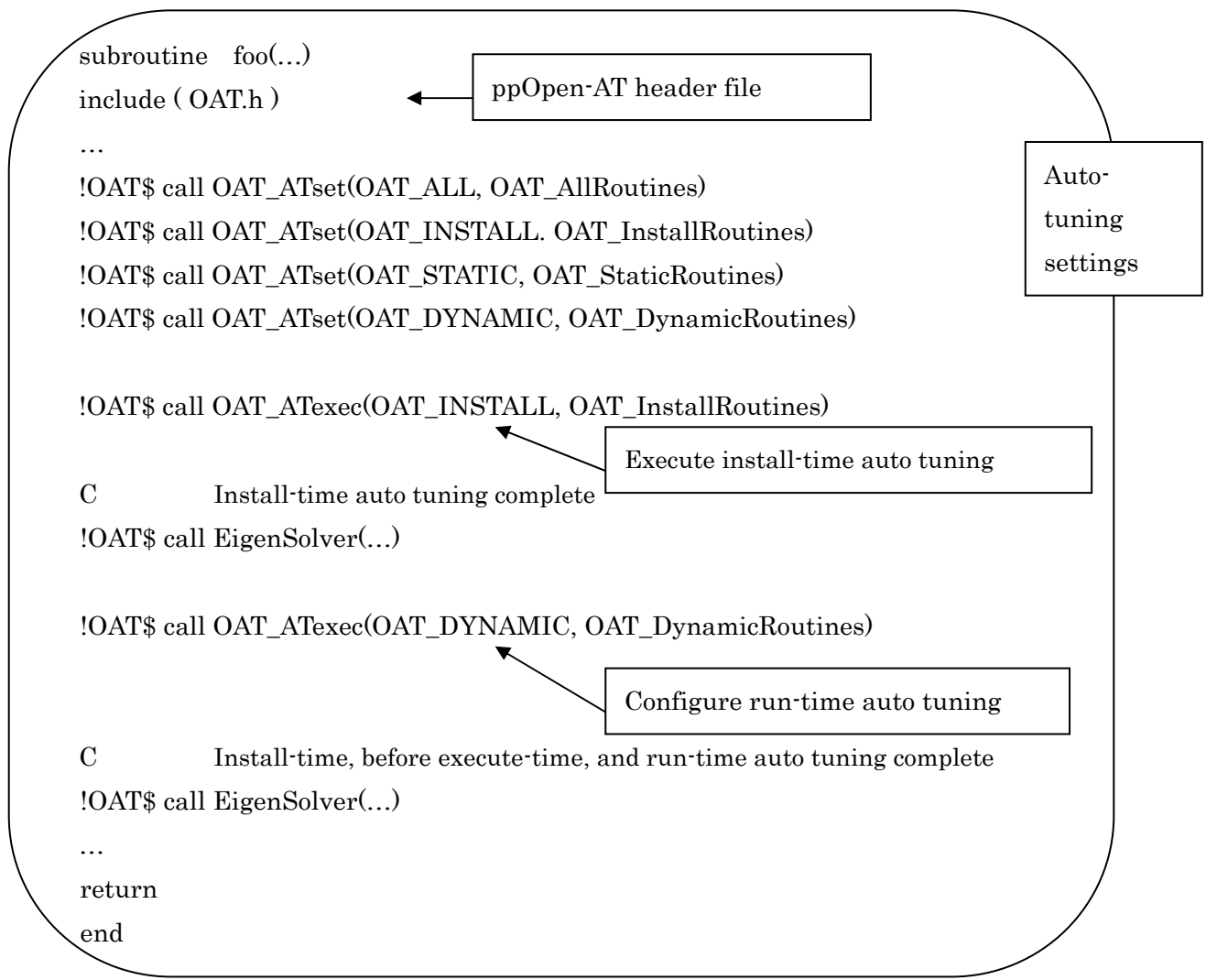

```
subroutine    foo(...)
include ( OAT.h )                    ppOpen-AT header file
...
!OAT$ call OAT_ATset(OAT_ALL, OAT_AllRoutines)
!OAT$ call OAT_ATset(OAT_INSTALL. OAT_InstallRoutines)         Auto-
!OAT$ call OAT_ATset(OAT_STATIC, OAT_StaticRoutines)          tuning
!OAT$ call OAT_ATset(OAT_DYNAMIC, OAT_DynamicRoutines)        settings

!OAT$ call OAT_ATexec(OAT_INSTALL, OAT_InstallRoutines)
                                          Execute install-time auto tuning
C          Install-time auto tuning complete
!OAT$ call EigenSolver(...)

!OAT$ call OAT_ATexec(OAT_DYNAMIC, OAT_DynamicRoutines)
                                          Configure run-time auto tuning
C          Install-time, before execute-time, and run-time auto tuning complete
!OAT$ call EigenSolver(...)
...
return
end
```

In this scenario, an end user is utilizing the EigenSolver feature of the library provided by the developer. Before execute-time auto-tuning can be performed, the following sequence must be completed using the ppOpen-AT interface within the user's subroutine, "pooh".

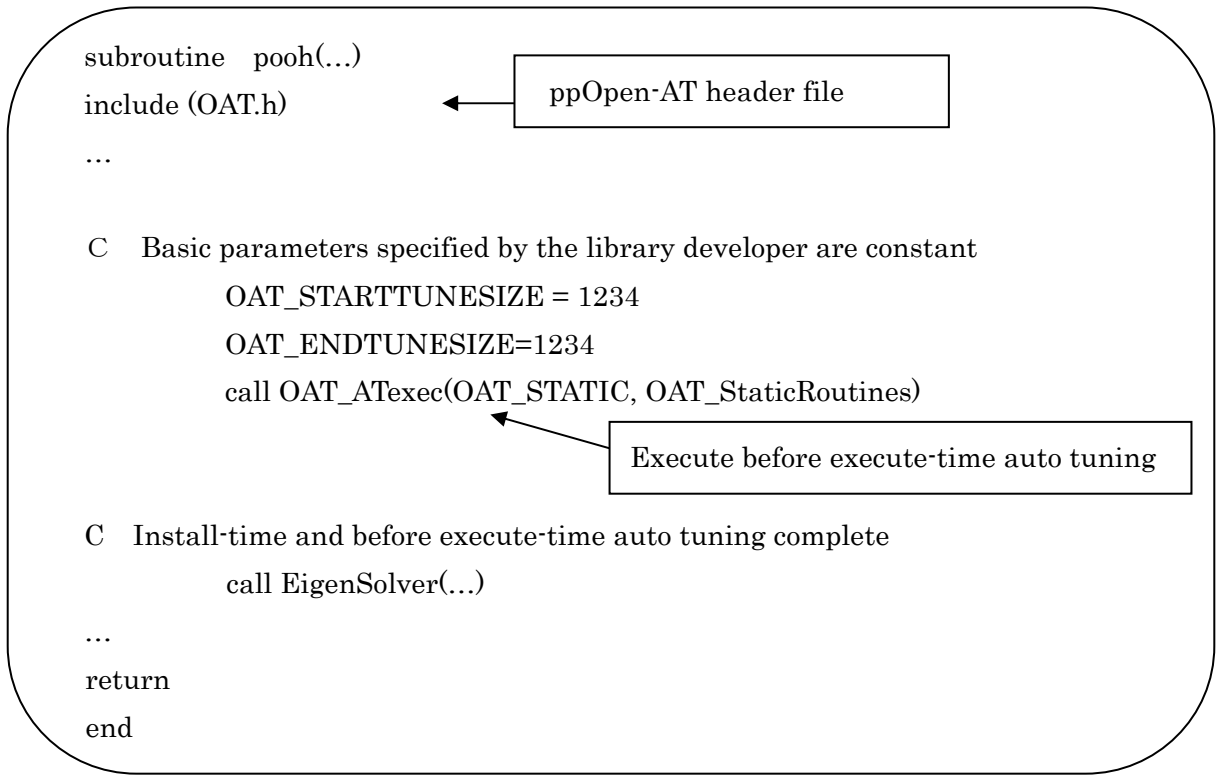

### 4.2.1 Sample Use of Install-time Auto Tuning Routines

Install-time auto-tuning routines are typically designed to be executed only once after the library installation. As a result, they are not executed again during subsequent invocations by default. If you wish to process them again, you must initialize them by calling the OAT_ATInstallInit initialization interface described below.

**OAT_ATInstallInit** ( OAT_InstallRoutines )

The OAT_ATInstallInit procedure undoes the tuning on the install-time tuning region specified in OAT_InstallRoutines.

**Sample Usage**

!OAT$ call OAT_ATInstallInit(OAT_InstallRoutines)

We assume that you want to run install-time auto tuning again.

In addition to the parameter inference performed upon installation as shown in **Sample Program 1**, install-time auto tuning routines are meant to measure those parameters that can be measured during installation. This specification is written using the following define function.

**Sample Program 2**

Parameter determination using the define function in an install-time routine

```
!OAT$   install   define  ( CacheSize,   CacheLine )   region   start
!OAT$   name   SetCacheParam
!OAT$   parameter   (out CacheSize,   out CacheLine )
…
CacheSize =….
CacheLine=….
!OAT$   install   define  ( CacheSize,   CacheLine )   region   end
```

The install-time routine saves these parameters specification values in a **parameter information file**. The file name is OAT_InstallParam.dat, and it is stored in text format.

Example: Contents of OAT_InstallParam.dat parameter information file
```
(SetCacheParam
  (CacheSize    64)
  (CacheLine    8)
 )
```

## 4.2.2 Sample Use of Before execute-time Auto Tuning Routine

Before execute-time auto tuning determines parameters based on detailed information before the end user executes the library. Before execute-time auto tuning tunes parameters on the condition that the end user guarantees the values of the basic parameter described in section 2.4.

The user specifies the parameters to be provided before library execution in the OAT_StaticParamDef.dat parameter information file. Basic parameters can be configured via substitution statements in the program. Here, **basic parameters** are required for install-time and before execute-time auto tuning. By default, they are referred to internally using parameter names specified by the system.

The default basic-information parameter settings are shown below.

**List of Default Basic Parameters**

<Default Basic Parameters>::=
   ( **OAT_NUMPROCS** | **OAT_STARTTUNESIZE** | **OAT_ENDTUNESIZE** | **OAT_SAMPDIST** )
      **OAT_NUMPROCS**:   Integer.   Indicates the number of processors to use.
      **OAT_STARTTUNESIZE**:   Integer. The starting problem size for
         auto tuning with regard to the default basic parameters.
      **OAT_ENDTUNESIZE**:   Integer. The ending problem size for
         auto tuning with regard to the default basic parameters.
      **OAT_SAMPDIST**: Integer. The size of the increment used to increase
         the problem size for auto tuning with regard to the
         default basic parameters.

Note that before execute-time auto tuning will not run if the basic parameters are not set. Additionally, install-time auto tuning will not run unless OAT_NUMPROCS, OAT_STARTTUNESIZE, OAT_ENDTUNESIZE, and OAT_SAMPDIST are set.

Library developers wishing to set new basic parameters should use the OAT_BPset and OAT_BPsetName procedures below.

**OAT_BPset**( BPvalName )

The OAT_BPset procedure makes the basic parameter name specified by the parameter BPvalName into a new basic parameter.

**Sample Usage**
!OAT$ call OAT_BPset ( "nprocs" )

Makes the parameter nprocs into a basic parameter.

---
**OAT_BPsetName**( Kind, BPvalName, Name)

---

The OAT_BPsetName procedure sets the name of the information parameter relating to the basic parameter of the type specified by the parameter "Kind" to the name specified by the parameter Name, for the basic parameter specified in the parameter BPValName.

Here, the parameter Kind is as follows.

Kind ::=[ **STARTTUNESIZE** | **ENDTUNESIZE** | **SAMPDIST** ]

**STARTTTUNESIZE**: Sample start point information relating to
   basic parameter BPvalName

**ENDTUNESIZE**: Sample end point information relating to
   basic parameter BPvalName

**SAMPDIST**: Sample point interval information relating to
   basic parameter BPvalName

**Sample Usage**
!OAT$ call OAT_BPsetName("STARTTUNESIZE", "nprocs",
!OAT$ &   "OAT_NprocsStartSize")

Sets the sample start point parameter name for auto tuning for basic parameter nprocs to OAT_NprocsStartSize.

---
**OAT_BPsetCDF**( BPvalName, CDFKind)

---

The OAT_BPsetCDF procedure sets the method to be used for inferring non-sample points relating to the basic parameter name specified in the parameter BPvalName to the type of cost definition function specified by the parameter CDF Kind. Here, the parameter CDF Kind is as follows.

Kind ::=[ **least-squares** <order>| **user-defined** <mathematical expression>| **auto** ]

**least-squares** <order>: Specifies the parameter must be inferred by the least
   squares method via a polynomial expression. <order> Sets the order of the
   polynomial expression.

> **user-defined** <mathematical expression>: Infer using the least squares method, using the mathematical expression specified by the user
>
> **auto**: Allow the system to infer the parameter.

Please note that, by default, the cost definition function specified for the relevant tuning region is used without modification.

**Sample Usage**

!OAT$ call OAT_BPsetCDF("nprocs", "least-squares 5")

Infer parameters for basic parameter nprocs using the least squares method, as a fifth-order polynomial equation.

**Sample Program 3**

Specify the basic-parameter sample points when performing before execute-time tuning for 1,024, 2,048, and 3,072 dimensional parameters using 4 processors.

```
!OAT$   OAT_TUNESTATIC = 1
!OAT$   OAT_NUMPROCS = 4
!OAT$   OAT_STARTTUNESIZE = 1024
!OAT$   OAT_ENDTUNESIZE = 3072
!OAT$   OAT_SAMPDIST = 1024
!OAT$   call OAT_ATexec(OAT_STATIC, OAT_StaticRoutines)
```

Another way of specifying this is to write the following in the OAT_StaticParam_Def.dat file.

```
(BasicParam
    (OAT_TUNESTATIC   1)
    (OAT_NUMPROCS   4)
    (OAT_STARTTUNESIZE   1024)
    (OAT_ENDTUNESIZE   3072)
    (OAT_SAMPDIST   1024)
)
```

It is expected that the unroll feature will be used in before execute-time routines.

Below is a sample of this usage.

**Sample Program 4a**

This before execute-time routine determines the optimum parameter, unrolling the loop in the program below to 16 levels. Parameter measurement is performed at each of levels 1 to 16 (i.e. exhaustive search). Note, however, that it is assumed that the sample points for the basic parameter have been set as in **Sample Program 3**. Since the only loop entry variable is $n$, parameter $n$ is interpreted as the default basic parameter.

```
!OAT$   static   unroll   (i, j)   region   start
!OAT$   name   MyMatMul
!OAT$   varied   (i, j)   from   1   to   16
do i=1, n
   do j=1, n
      do k=1, n
         A(i, j) = A(i, j) +B(i, k)*C(k, j)
      enddo
   enddo
enddo
!OAT$   static   unroll   (i, j)   region   end
```

In the case of this example, the optimum output parameters (i.e. the parameters for which processing is fastest) are the values i, j.

The before execute-time auto tuning routine writes this optimized parameter to the OAT_StaticParam.dat parameter information file.

Example: Contents of OAT_StaticParam.dat parameter information file:
(MyMatMul
    (OAT_NUMPROCS   4)
    (OAT_SAMPDIST    1024)
    (OAT_PROBSIZE   1024
       (MyMatMul_I    4)
       (MyMatMul_J    8)   )
    (OAT_PROBSIZE   2048
       (MyMatMul_I    4)

```
   (MyMatMul_J    9)   )
  (OAT_PROBSIZE   3072
     (MyMatMul_I    5)
     (MyMatMul_J    10)   )
)
```

Before execute-time auto-tuning routine parameter determination can be performed only by using parameters specified in the program, by calling the parameters of the install-time auto-tuning routine (the data in the parameter information file). **Sample Program 5** shows an example of this.

**Sample Program 4b**

This before execute-time routine determines the optimum parameter, unrolling the loop in the program below to 16 levels. Parameter measurement is performed at each of levels 1 to 16 (i.e. exhaustive search). Note, however, that it is assumed that the basic information parameters have been specified as in **Sample Program 3**.

In the example below, because *n* is not the only loop-termination variable (the variable *nprocs* also exists), it is not known whether the variable *n* or *nprocs* is the default basic parameter. Consequently, the user must specify the basic parameter via the parameter subtype specifier.

```
!OAT$   static   unroll   (i, j)   region   start
!OAT$   name   MyMatMul
!OAT$   parameter(bp n)
!OAT$   varied   (i, j)   from   1   to   16
do i=1, n/nprocs
  do j=1, n
     do k=1, n
        A(i, j) = A(i, j) +B(i, k)*C(k, j)
     enddo
  enddo
enddo
!OAT$   static   unroll   (i, j)   region   end
```

**Sample Program 4c**

This before execute-time routine determines the optimum parameter, unrolling the loop in the program below to 16 levels. Parameter measurement is performed at each of levels 1 to 16 (i.e. exhaustive search). Note, however, that it is assumed that the basic information parameters have been specified as in **Sample Program 3**. In the example below, both *n* and *nprocs* are specified as basic parameters.

```
!OAT$ call OAT_BPsetVal("nprocs")
!OAT$ call OAT_BPsetName(STARTTUNESIZE, "nprocs",
!OAT$ &          "OAT_NprocsStartSize")
!OAT$ call OAT_BPsetName(ENDTUNESIZE, "nprocs",
!OAT$ &                "OAT_NprocsEndSize")
!OAT$ call OAT_BPsetName(SAMPDIST, "nprocs",
!OAT$ &              "OAT_NprocsSampDist")
!OAT$ OAT_NprocsStartSize = 1
!OAT$ OAT_NprocsEndSize = 8
!OAT$ OAT_NprocsSampDist = 1
!OAT$ static   unroll   (i, j)   region   start
!OAT$ name    MyMatMu
!OAT$ parameter(bp n, bp nprocs)
!OAT$ varied   (i, j)   from   1   to   16
do i=1, n/nprocs
   do j=1, n
      do k=1, n
          A(i, j) = A(i, j) +B(i, k)*C(k, j)
      enddo
   enddo
enddo
!OAT$ static   unroll   (i, j)   region   end
```

**Sample Program 5**

Determine the optimum implementation method referring to the parameter CacheSize of the install-time auto-tuning routine, and using the basic parameter information for problem size and number of processors set by the end user before execution. Here, the selection is made automatically, using a given standard (in this case, the run-time estimate provided by the user)[§].

```
!OAT$   static  select  region  start
!OAT$   name  ATfromCacheSize
!OAT$   parameter  (in CacheSize,  in OAT_PROBSIZE,
!OAT$ &             in OAT_NUMPROC)
!OAT$   select  sub  region  start
!OAT$   according  estimated
!OAT$ &        2.0d0*CacheSize*OAT_PROBSIZE*OAT_PROBSIZE
!OAT$ &        / (3.0d0*OAT_NUMPROC)
                Target process 1
!OAT$   select  sub  region  end
!OAT$   select  sub  region  start
!OAT$   according  estimated  4.0d0*CacheSize*OAT_PROBSIZE
!OAT$ &        *dlog(OAT_PROBSIZE) / (2.0d0*OAT_NUMPROC)
                Target process 2
!OAT$   select  sub  region  end
!OAT$   static  select  region  end
```

### 4.2.3 Sample Use of a Run-time Auto Tuning Routine

Run-time auto tuning routines determine performance parameters based on information that can be obtained at run-time. When determining these performance parameters, it is possible to reference parameters specified in the program, parameters determined by an install-time auto-tuning routine, and parameters specified by a before execute-time auto-tuning routine.

---

[§] In this case, the selection is based on execution time. Note that execution time is generally not the only standard for AT region selection. For example, subroutine cost is also a possible selection standard. This type of selection standard can be encoded using the select specifier.

Below is a description of the select function, the use of which can be expected in run-time routines.

**Sample Program 6**

Select the optimum process based on the parameters *eps* and *iter* defined in the program. Specifically, set condition *iter* to less than 5, in order to make *eps* the minimum value.

```
!OAT$  dynamic  select  ( eps,  iter )  region  start
!OAT$  name  PricondSelect
!OAT$  parameter  ( in eps,  in iter )
!OAT$  according  min ( eps )  .and.  condition ( iter < 5 )
!OAT$    select  sub  region  start
      Target process 1
       eps=…
!OAT$    select  sub  region  end
!OAT$    select  sub  region  start
      Target process 2
     eps=…
!OAT$    select  sub  region  end
!OAT$  dynamic  select  ( eps,  iter )  region  end
```

**OAT_DynPefThis**( Name )

The OAT_DynPefThis procedure specifies that the optimization of the tuning region name performed by the run-time auto-tuning routine specified in the parameter "Name" is to be performed at the location written in this procedure. Note that when using this procedure, the execution of ((auto) tuning for the tuning region name is performed using optimized parameters. In other words, no parameter tuning is performed by the tuning region specified by Name.

**Sample Usage**

!OAT$ call OAT_DynPefThis ( "DMatVec" )

Perform the auto tuning of the auto tuning region DMatVec here.

**Sample Program 7**

Regarding the determination of the optimum number of loop-unrolling levels, re-use the parameters optimized by run-time optimization, and execute the process optimized in the AT region in question.

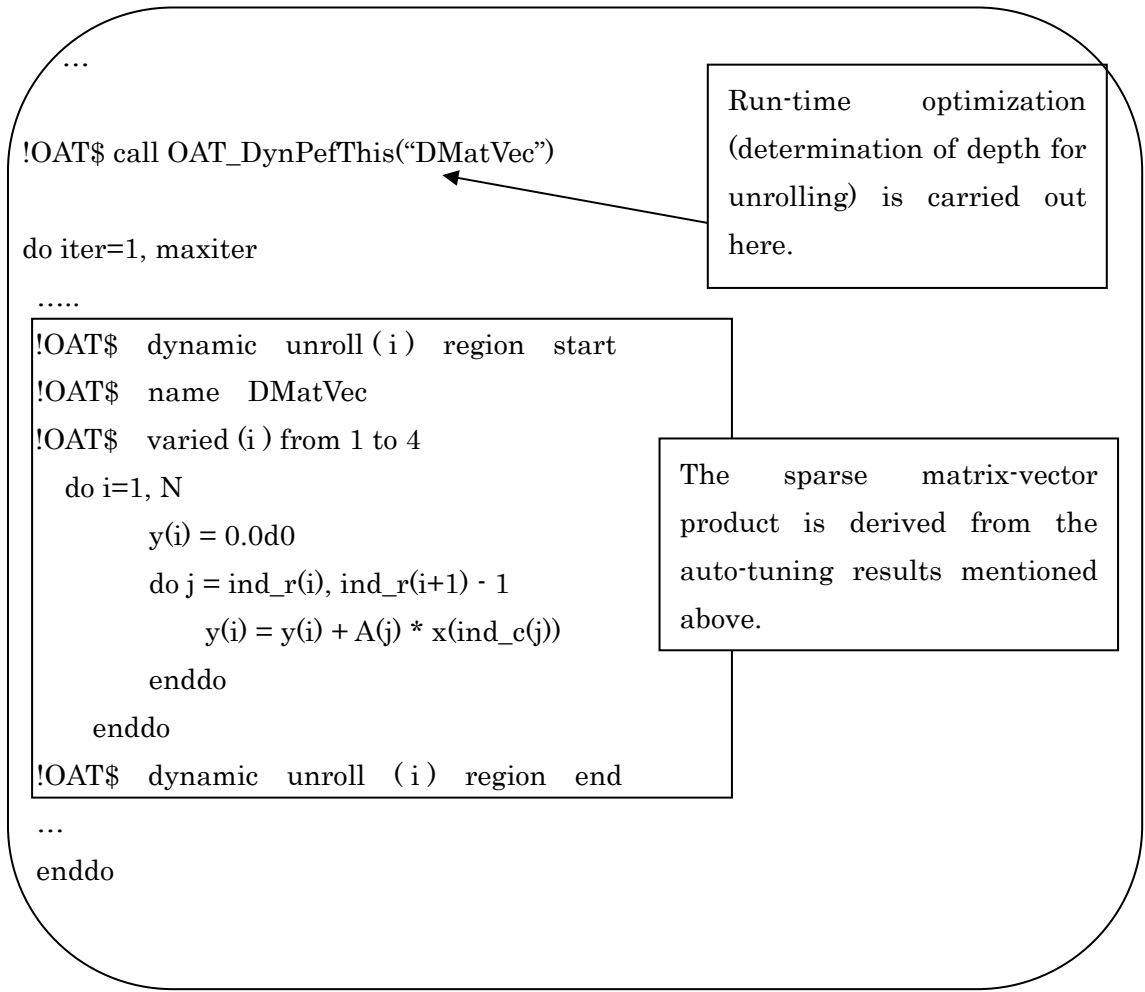

```
...
!OAT$ call OAT_DynPefThis("DMatVec")

do iter=1, maxiter
  .....
!OAT$   dynamic   unroll ( i ) region   start
!OAT$   name   DMatVec
!OAT$   varied ( i ) from 1 to 4
    do i=1, N
          y(i) = 0.0d0
          do j = ind_r(i), ind_r(i+1) - 1
                y(i) = y(i) + A(j) * x(ind_c(j))
          enddo
      enddo
!OAT$   dynamic   unroll ( i ) region   end
...
enddo
```

Run-time optimization (determination of depth for unrolling) is carried out here.

The sparse matrix-vector product is derived from the auto-tuning results mentioned above.

## 4.3 Sample Use of the Auto Tuning Code Generation Command

To generate actual parallel-computing library code with auto-tuning features (Fortran90 + MPI), do the following:

```
>OATCodeGen   test.f
```

Here we assume that test.f is a program including ppOpen-AT directives.

After executing the command above, an OAT directory is created beneath the current directory, in which the following files are created:

> ./OAT/OAT_test.f:
>
>     Fortran90 + MPI source code with embedded auto-tuning code in test.f
>
> **./OAT/OAT_InstallRoutines.f**:
>
> Subroutines extracted from test.f that perform install-time auto tuning
>
> **./OAT/OAT_StaticRoutines.f**:
>
> Subroutines extracted from test.f that perform before execute-time auto tuning
>
> **./OAT/OAT_DynamicRoutines.f**:
>
>     Subroutines extracted from test.f that perform run-time auto tuning
>
> **./OAT/OAT_ControlRoutines.f**:
>
>     Subroutines that control auto-tuning code

If these files already exist when the preprocessor starts, non-overlapping source code and subroutines are added to the above files.

## 4.3.1 Run-time Options

OATCodeGen can specify the following run-time options.

> **－debug**
>
>   **OFF**: Do not generate debugging code
>
>   **ON**: Generate code debugging code at debug level x, specified
>
>       by the OAT_PRINT parameter.
>
> **－visualization**
>
>   **OFF**: Do not output an auto-tuning trace file (default value)
>
>   **ON**: Output an auto-tuning trace file, and use it to create a visualization

The name of the auto-tuning trace file created when －visualization is set to ON is:

OATATlog.dat

**Sample Usage**

> OAT -debug ON -visualization ON   test.f

Specify the generation of debugging code, and the output of an auto-tuning trace file of the trace of auto-tuning mode for program test.f, containing ppOpen-AT specifiers. This auto-tuning trace file can be used to dynamically view tuning progress in a visualizer (VizOAT).

# 5.　Extended Functions

## 5.1 Overview

In versions after version 0.2, or in other later versions, new functions for loop transformation are available. The main functions are loop split and loop fusion with data dependences. In addition to that, re-ordering of sentences in a loop is also available. In this session, we explain two key examples for the new extended functions to the original functions of ppOpen-AT.

## 5.2 Functions for loop split and fusion with data dependences

The ppOpen-AT functions for loop split and loop fusion are explained with the following Sample Program 8.

### Sample Program 8

The following is an example of how to adapt the loop split and fusion functions to the heaviest kernel in ppOpen-APPL/FDM. There is a flow dependency between the definition of QG and several uses of QG. Hence in general, it is difficult to perform loop splitting using compilers.

The new directive for specifying loop split and loop fusion is "!oat\$ install LoopFusionSplit region start" ~ "!oat\$ install LoopFusionSplit region end".

To define a loop split point, "!oat\$ SplitPoint　(K, J, I)" can be used. The target loops, including loop induction variables, for splitting can be defined with the directive. In this case, loops for K, J, and I can be adapted for the loop splitting operation.

The re-computation sentences when a loop split is done can specified with "!oat\$ SplitPointCopyDef　region　start" ~ "!oat\$ SplitPointCopyDef　region　end". Points to which to copy the sentences that are defined by "!oat\$ SplitPointCopyDef region　start" ~ "!oat\$ SplitPointCopyDef　region　end" can be defined with "!oat\$ SplitPointCopyInsert".

```
!oat$ install LoopFusionSplit region start          ← Specify loop split & loop fusion
!$omp parallel do private (k, j, i, STMP1, STMP2, STMP3, STMP4, RL, RM, RM2,
      RMAXY, RMAXZ, RMAYZ, RLTHETA, QG )
DO K = 1, NZ
DO J = 1, NY
DO I = 1, NX
      RL   = LAM (I,J,K);    RM  = RIG (I,J,K);    RM2 = RM + RM
      RLTHETA   = (DXVX(I,J,K)+DYVY(I,J,K)+DZVZ(I,J,K))*RL
!oat$ SplitPointCopyDef   region   start            ← Define re-computation sentences
      QG   = ABSX(I)*ABSY(J)*ABSZ(K)*Q(I,J,K)       ← Define data with sentence
!oat$ SplitPointCopyDef   region   end
      SXX (I,J,K) = ( SXX (I,J,K) + (RLTHETA + RM2*DXVX(I,J,K))*DT )*QG
      SYY (I,J,K) = ( SYY (I,J,K) + (RLTHETA + RM2*DYVY(I,J,K))*DT )*QG    ← Use the data
      SZZ (I,J,K) = ( SZZ (I,J,K) + (RLTHETA + RM2*DZVZ(I,J,K))*DT )*QG
!oat$ SplitPoint   (K, J, I)                        ← The point for the loop split with the loop

      STMP1 = 1.0/RIG(I,J,K);   STMP2 = 1.0/RIG(I+1,J,K);   STMP4 = 1.0/RIG(I,J,K+1)
      STMP3 = STMP1 + STMP2
      RMAXY = 4.0/(STMP3 + 1.0/RIG(I,J+1,K) + 1.0/RIG(I+1,J+1,K))
      RMAXZ = 4.0/(STMP3 + STMP4 + 1.0/RIG(I+1,J,K+1))
      RMAYZ = 4.0/(STMP3 + STMP4 + 1.0/RIG(I,J+1,K+1))      ← Use the data
!oat$ SplitPointCopyInsert         ← Point for copy
      SXY (I,J,K) = ( SXY (I,J,K) + (RMAXY*(DXVY(I,J,K)+DYVX(I,J,K))*DT )*QG
      SXZ (I,J,K) = ( SXZ (I,J,K) + (RMAXZ*(DXVZ(I,J,K)+DZVX(I,J,K))*DT )*QG
      SYZ (I,J,K) = ( SYZ (I,J,K) + (RMAYZ*(DYVZ(I,J,K)+DZVY(I,J,K))*DT )*QG
END DO
END DO
END DO                      ← Specify loop split & fusion (region end)
!$omp end parallel do
!oat$ install LoopFusionSplit region end
```

By adapting the preprocessor of ppOpen-AT, we can obtain candidates for code that provide all combinations by adapting loop split and loop fusion by using the above directives. In the case of sample 8, we can obtain the following 8 candidates:

1. #1 [Baseline] : Original three-nested loop.
2. #2 [Spilt] : Loop split for the k-loop (separated into two three-nested loops).
3. #3 [Split] : Loop split for the j-loop.
4. #4 [Split] : Loop split for the i-loop.
5. #5 [Fusion] : Loop fusion for the k-loop and j-loop (a two-nested loop).
6. #6 [Split and Fusion] : Loop fusions for the k-loop and j-loop for the loops in #2.
7. #7 [Fusion] : Loop fusions for the k-loop, j-loop, and i-loop (loop collapse).
8. #8 [Split and Fusion] : Loop fusions for the k-loop, j-loop, and i-loop for the loops in #2 (the loop collapses for the two separated loops).

## 5.3 Function for re-ordering of sentences in a loop

The ppOpen-AT function for re-ordering of sentences is explained in the following Sample Program 9.

### Sample Program 9

The following is an example of how to adapt re-ordering of sentences in loops. In this example, the functions of loop split and fusion to one of the heavy kernels in ppOpen-APPL/FVM is applied. In this case, loop fusion is specified, but you can also specify other loop transformations, such as a loop split only.

The target sentences for the re-ordering of sentences can be specified with "!OAT$ RotationOrder sub region start "~ "!OAT$ RotationOrder sub region end ". This specification is needed for two regions. Each sentence that is specified the directive is re-ordered.

```
!OAT$ install LoopFusion region start          ┌─ Specify the loop fusion in this example. ─┐
do k = NZ00, NZ01
    do j = NY00, NY01
        do i = NX00, NX01              ┌─ Specify re-ordering of sentences (1). ─┐
    !OAT$ RotationOrder sub region start
            ROX = 2.0_PN/( DEN(I,J,K) + DEN(I+1,J,K) )
            ROY = 2.0_PN/( DEN(I,J,K) + DEN(I,J+1,K) )
            ROZ = 2.0_PN/( DEN(I,J,K) + DEN(I,J,K+1) )
    !OAT$ RotationOrder sub region end
    !OAT$ RotationOrder sub region start        ┌─ Specify re-ordering of sentences (2). ─┐
            VX(I,J,K) =
                VX(I,J,K) + ( DXSXX(I,J,K)+DYSXY(I,J,K)+DZSXZ(I,J,K) )*ROX*DT
            VY(I,J,K) =
                VY(I,J,K) + ( DXSXY(I,J,K)+DYSYY(I,J,K)+DZSYZ(I,J,K) )*ROY*DT
            VZ(I,J,K) =
                VZ(I,J,K) + ( DXSXZ(I,J,K)+DYSYZ(I,J,K)+DZSZZ(I,J,K) )*ROZ*DT
    !OAT$ RotationOrder sub region end
                end do
            end do
        end do
!OAT$ install LoopFusion region end
```

One of code segments generated for the original 3-nested loop is as follows.

```
do k = NZ00, NZ01
    do j = NY00, NY01
        do i = NX00, NX01
        ROX = 2.0_PN/( DEN(I,J,K) + DEN(I+1,J,K) )
        VX(I,J,K) = VX(I,J,K) +DXSXX(I,J,K)+DYSXY(I,J,K)+DZSXZ(I,J,K) )*ROX*DT
        ROY = 2.0_PN/( DEN(I,J,K) + DEN(I,J+1,K) )
        VY(I,J,K) = VY(I,J,K) + ( DXSXY(I,J,K)+DYSYY(I,J,K)+DZSYZ(I,J,K) )*ROY*DT
        ROZ = 2.0_PN/( DEN(I,J,K) + DEN(I,J,K+1) )
        VZ(I,J,K) = VZ(I,J,K) + ( DXSXZ(I,J,K)+DYSYZ(I,J,K)+DZSZZ(I,J,K) )*ROZ*DT
        end do
```

end do
    end do

# 6. ppOpen-AT Internal Specifications

## 6.1 System Parameters

The following is a list of system parameters maintained by ppOpen-AT. Please note that these parameters are reserved words and cannot be defined by users.

**List of System Parameters (Reserved Words)**

*Tuning Type Specifiers*

| |
|---|
| **OAT_ALL**: An integer type with value 0. Indicates all tuning types (install-time, before execute-time, and run-time) |
| **OAT_INSTALL**: An integer type with value 1. Indicates install-time optimization. |
| **OAT_STATIC**: An integer type with value 2. Indicates before execute-time optimization. |
| **OAT_DYNAMIC**: An integer type with value 3. Indicates run-time optimization. |

*Tuning Region Name Storage*

| |
|---|
| **OAT_AllRoutines**: A string type. All tuning region names are stored. |
| **OAT_InstallRoutines**: A string type. The names of tuning regions for which install-time auto tuning is to be performed are stored. |
| **OAT_StaticRoutines**: A string type. The names of tuning regions for which before execute-time auto tuning is to be performed are stored. |
| **OAT_DynamicRoutines**: A string type. The names of tuning regions for which run-time auto tuning is to be performed are stored. |

*Default Basic Parameters*

| |
|---|
| **OAT_NUMPROCS**: Integer. Holds the number of processors to be used. |
| **OAT_STARTTUNESIZE**: Integer. Holds the starting sample point value for the default basic parameters. |
| **OAT_ENDTUNESIZE**: Integer. Holds the ending sample point value for the default basic parameters. |
| **OAT_SAMPDIST**: Integer. Holds the sample point interval (increment) value for the default basic parameters. |

*System Control*

---

**OAT_TUNESTATIC**: Boolean. Specifies whether to execute before execute-time
auto tuning.

    **.true.**: Execute. **.false.**: Do not execute.

**OAT_TUNEDYNAMIC**: Boolean. Specifies whether to execute run-time
auto tuning.

    **.true.**: Execute. **.false.**: Do not execute.

**OAT_DEBUG**: Integer. Specifies the debug print level.

    **0**: None. **Value x of 1 or greater**: Debug printing at level x.

---

## 6.2 Input and Output Files

This section describes the input and output files (parameter information files)
handled by ppOpen-AT. There are two types of I/O file: Files automatically generated by
ppOpen-AT (**system specification files**) and files specified by the user for debugging and
the like (**user specification files**).

**System specification files:**

---

**OAT_InstallParam*X*.dat**: For install-time auto tuning routine parameter output

**OAT_StaticParam*X*.dat**: For before execute-time auto tuning routine
parameter output

---

**User specification files**

---

**OAT_InstallParamDef*X*.dat**: For install-time auto tuning routine parameter
specification

**OAT_StaticParamDef*X*.dat**: For before execute-time auto tuning routine
parameter specification

**OAT_DynamicParamDef*X*.dat**: Input file for before execute-time auto tuning routine
parameter specification

---

Note: *X* holds the name of the AT region in question.

## 6.2.1 Input Files

Input files consist of both user specification files and system specification files.

**User specification files**

    **OAT_InstallParamDef*X*.dat**: For install-time auto tuning routine parameter
specification

**OAT_StaticParamDef*X*.dat**: For before execute-time auto tuning routine
   parameter specification

**OAT_DynamicParamDef*X*.dat**: For run-time auto tuning routine
   parameter specification

**System specification files:**

   install-time routines:   None

   Before execute-time routines:

      **OAT_InstallParam*X*.dat**

   Run-time routines:

      **OAT_StaticParam*X*.dat,**

      **OAT_DynamicParam*X*.dat**

## 6.2.2 Output Files

   Output files consist of system specification files only.

**System specification files:**

   install-time routines:

      **OAT_InstallParam*X*.dat**

   Before execute-time routines:

      **OAT_StaticParam*X*.dat**

   Run-time routines:

      **OAT_DynamicParam*X*.dat**

## 6.2.3 Input / Output File Format

   The format of user specification files and system specification files is as follows.

```
<format>::=
(<name>
    (<key> <value>)
    [(<key > < value >)]
      ...
    )
    [<format>];
 < key >::= (<format> | parameter name) ;
 < value >::=[parameter value];
```

<name> specifies a tuning region name or basic parameter name. To specify a basic parameter, write BasicParam.

## 6.3 Collisions with Parameters in User Specification Files

When auto tuning is performed on a parameter specified in a user specification file (when an attempt is made to determine the parameter), this is called a **parameter collision**.

When there is a parameter collision, auto tuning halts, and the user-specified parameter is forcibly set.

When the system detects a parameter collision, it assumes that the user wants to halt the auto-tuning feature in order to debug or the like. To put it the other way around, the user can perform debugging by defining parameter information in a user specification file.

## 6.4 Nesting of Statements

This section defines specifier nesting.

### 6.4.1 Nesting Availability and Depth

The type of nesting available in auto-tuning is defined in Table 1.

Table 1. Availability of Nesting by Auto Tuning Type

| Nesting part (superior part) | Nested part (subordinate part) | | |
|---|---|---|---|
| | install | static | dynamic |
| install | yes | no | no |
| static | yes | yes | no |
| dynamic | yes | yes | yes |

Table 2 defines combinations of features that can be nested.

Table 2. Nesting Availability by Feature

| Nesting part (superior part) | Nested part (subordinate part) | | | |
|---|---|---|---|---|
| | define | variable | select | unroll |
| define | yes | yes | yes | yes |
| variable | yes | yes | yes | yes |
| select | yes | yes | yes | yes |
| unroll | no | no | no | no |

The maximum nesting depth (how far down elements can be nested) is currently as shown below. In other words, the maximum nesting depth is three.

> **Nesting depth = 3 or fewer**

## 6.4.2 Parameter Search Order (extended feature)

The method used for searching for nested parameters is determined by the method specified by the outermost tuning region. The parameter search method can be annotated as follows.

> subtype specifier ::=
> **(exhaustive search | AD-HOC method)**

Now, let us assume that there are *m* tuning regions with parameters P_i (i=1, 2, ... m), each needing to vary N_i parameters. In this case, the parameters are expressed as follows:

$$P = (V(P\_1), V(P\_2), ..., V(P\_m)),$$

where V(P_i) expresses 1 of the N_i parameters of parameter P_i.

**Exhaustive search:**

> **!OAT\$   search   Brute-force**

In an exhaustive search, all combinations are investigated. In other words, under this method all combinations of parameter P are searched.

Consequently, the number of parameter combinations is $\Pi N\_i$.

**AD-HOC method:**

| !OAT$　search　AD-HOC |
| --- |

When the AD-HOC method is used, not all combinations of parameter P are searched.

The search starts with a given parameter with a set initial value, a varied P_m, and the optimum parameter found and set. Next, it is P_m-1, then the optimum parameter is found and set, and so on. The algorithm then repeats the process until P_1.

Consequently, the number of parameter combinations is $\Sigma N\_i$.

**Action When Different Search Methods are Specified for Different Nested Specifiers**

In general, the search begins from the innermost AT region, and is made to match the outermost search method. However, if the inner method is AD-HOC, and the outer method is exhaustive, it will be treated as if the parameters of the AD-HOC specified AT regions are constant values.

## Sample Program 10

How is parameter searching carried out for the following nested processes?

```
!OAT$  static  variable  (BL)  region  start
!OAT$  name  ABlockRoutine
!OAT$  varied  1  from  16
do iter=1,  n,  BL
...
!OAT$  static  unroll  ( i, j )  region  start
!OAT$  name  Kernel1
!OAT$  varied ( i, j )  from  1  to  32
  do i=1+iter, n
   do j=1+iter, n
     do k=1+iter, n
       .....
     enddo
    enddo
  enddo
!OAT$  static  unroll ( i, j )  region  end
....
!OAT$  static  unroll (i, j)  region  start
!OAT$  name  Kernel2
!OAT$  varied  ( l, m )  from  1  to  32
  do l=1+iter, n
    do m=1+iter, n
      do p=1+iter, n
        .....
      enddo
    enddo
  enddo
!OAT$  static  unroll ( l, m )  region  end
...
enddo
!OAT$  static  variable  (BL)  region  end
```

Here, we assume that the parameter ordering is (BL, (i,j),(l,m)).

In the case of the above example, an exhaustive search is performed for all tuning regions: AblockRoutine, Kernel1, and Kernel2. Here, the parameter search proceeds as follows:

(1,(1,1),(1,1)), (1,(1,1),(1,2)),....,(1,(1,1),(1,32)),

(1,(1,1),(2,1)), (1,(1,1),(2,2)),...,(1,(1,1),(2,32)),

...,

(1,(1,2),(1,1)),(1,(1,2),(1,2)),...,(1,(1,2),(1,32)), ....,

Thus, there are 16*32*32*32*32 = 1,677,216 searches.

Let us assume that in the above example, the method for all tuning regions (AblockRoutine, Kernel1, and Kernel2) is AD-HOC. In this case, the search will be as follows:

(1,(1,1),(1,1)),(1,(1,1),(1,2)),....,(1,(1,1),(1,32)): Fastest parameter is determined (e.g. 8)

(1,(1,1),(1,8)),(1,(1,1),(2,8)),....,(1,(1,1),(32,8)): Fastest parameter is determined (e.g. 4)

(1,(1,1),(4,8)),(1,(1,2),(4,8)),....,(1,(1,32),(4,8)): Fastest parameter is determined (e.g. 5)

(1,(1,5),(4,8)),(1,(2,5),(4,8)),...

In other words, there are 16+32+32+32+32 = 144 parameter searches.

Let us assume that in the above example, the method for tuning region AblockRoutine is exhaustive search, and that for tuning regions Kernel1 and Kernel2, it is AD-HOC. In this case, the search will be as follows:

(1,(1,1),(1,1)),(1,(1,1),(1,2)),....,(1,(1,1),(1,32)): Fastest parameter is determined (e.g. 8)

(1,(1,1),(1,8)),(1,(1,1),(2,8)),....,(1,(1,1),(32,8)): Fastest parameter is determined (e.g. 4)

(1,(1,1),(4,8)),(1,(1,2),(4,8)),....,(1,(1,32),(4,8)): Fastest parameter is determined (e.g. 5)

(1,(1,5),(4,8)),(1,(2,5),(4,8)),....,(1,(32,5),(4,8)): Fastest parameter is determined (e.g. 6)

(1,(6,5),(4,8)),(2,(6,5),(4,8)),...,

In other words, there are 16+32+32+32+32 = 144 parameter searches.

Now let us assume that in the above example, the method for tuning region AblockRoutine is AD-HOC, and that for tuning regions Kernel1 and Kernel2, it is

exhaustive search. In this case, the search will be as follows:

(1,(1,1),(1,1)),(1,(1,1),(1,2)),…,(1,(1,1),(1,32))                                                ,

(1,(1,1),(2,1)),(1,(1,1),(2,2)),…,(1,(1,1),(2,32)) ,

…

(1,(1,1),(32,1)),(1,(1,1),(32,2)),…,(1,(1,1),(32,32)): Fastest parameter is determined (e.g. (3,9))

(1,(1,1),(3,9)),(1,(1,2),(3,9)),…,(1,(1,32),(3,9)),

(1,(2,1),(3,9)),(1,(2,2),(3,9)),…,(1,(2,32),(3,9)),

…

(1,(32,1),(3,9)),(1,(32,2),(3,9)),…,(1,(32,32),(3,9)): Fastest parameter is determined (e.g. (2,8))

(1,(2,8),(3,9)),(2,(2,8),(3,9)),…,(16,(2,8),(3,9)): Fastest parameter is determined (e.g. 6)

In other words, there are 16+32*32+32*32 = 2,064 parameter searches.

Note that if no search method is specified, the default methods are as follows.

| |
|---|
| Feature: Default search method |
| **define**: None (no need for search) |
| **variable**: Exhaustive search |
| **select**: AD-HOC search |
| **unroll**: Exhaustive search |

# 7. Conclusion

This technical report outlines the specifications and usage of the ppOpen-AT auto-tuning processing directives for software developers. The auto-tuning directives encoded in ppOpen-AT represent the unique expertise of developers. By encapsulating the craftsmanship of software engineers in ppOpen-AT, this knowledge can be transmitted to other engineers through the source code.

Previously, such knowledge was typically acquired on an individual basis and was not easily shareable. This resulted in the expertise of individuals becoming a form of obscure, craft-based technical knowledge. Given this context, the author believes that clearly encoding knowledge in the source code using ppOpen-AT will create a significant ripple effect within the engineering community.